


\font\titlefont = cmr10 scaled\magstep 4
\font\sectionfont = cmr10
\font\littlefont = cmr5
\font\eightrm = cmr8

\def\sss{\scriptscriptstyle}

\newcount\tcflag
\tcflag = 0  

\ifnum\tcflag = 0 \magnification = 1200 \fi  

\global\baselineskip = 1.2\baselineskip
\global\parskip = 4pt plus 0.3pt
\global\abovedisplayskip = 18pt plus3pt minus9pt
\global\belowdisplayskip = 18pt plus3pt minus9pt
\global\abovedisplayshortskip = 6pt plus3pt
\global\belowdisplayshortskip = 6pt plus3pt

\def\barsoff{\overfullrule=0pt}


\def\endignore{}
\def\ignore #1\endignore{}

\newcount\dflag
\dflag = 0


\def\monthname{\ifcase\month
\or January \or February \or March \or April \or May \or June%
\or July \or August \or September \or October \or November %
\or December
\fi}

\newcount\dummy
\newcount\minute  
\newcount\hour
\newcount\localtime
\newcount\localday
\localtime = \time
\localday = \day

\def\advanceclock#1#2{ 
\dummy = #1
\multiply\dummy by 60
\advance\dummy by #2
\advance\localtime by \dummy
\ifnum\localtime > 1440 
\advance\localtime by -1440
\advance\localday by 1
\fi}

\def\settime{{\dummy = \localtime%
\divide\dummy by 60%
\hour = \dummy
\minute = \localtime%
\multiply\dummy by 60%
\advance\minute by -\dummy
\ifnum\minute < 10
\xdef\spacer{0} 
\else \xdef\spacer{}
\fi %
\ifnum\hour < 12
\xdef\ampm{a.m.} 
\else
\xdef\ampm{p.m.} 
\advance\hour by -12 %
\fi %
\ifnum\hour = 0 \hour = 12 \fi
\xdef\timestring{\number\hour : \spacer \number\minute%
\thinspace \ampm}}}



\def\endtitle{}
\def\title#1\endtitle{\vskip.5in\titlefont
\global\baselineskip = 2\baselineskip
#1\vskip.4in
\baselineskip = 0.5\baselineskip\rm}

\def\endauthors{}
\def\authors#1\endauthors{#1}

\def\endabstract{}
\def\abstract#1\endabstract{\vskip .3in%
\centerline{\sectionfont\bf Abstract}%
\vskip .1in
\noindent#1}

\def\nopageonenumber{\footline={\ifnum\pageno<2\hfil\else
\hss\tenrm\folio\hss\fi}}  

\newcount\nsection
\newcount\nsubsection

\def\section#1{\global\advance\nsection by 1
\nsubsection=0
\bigskip\noindent\centerline{\sectionfont \bf \number\nsection.\ #1}
\bigskip\rm\nobreak}

\def\subsection#1{\global\advance\nsubsection by 1
\bigskip\noindent\sectionfont \sl \number\nsection.\number\nsubsection)\
#1\bigskip\rm\nobreak}


\def\appendix#1#2{\bigskip\noindent%
\centerline{\sectionfont \bf Appendix #1.\ #2}
\bigskip\rm\nobreak}


\newcount\nref
\global\nref = 1

\def\therefs{}


\def\ref#1#2{\xdef #1{[\number\nref]}
\ifnum\nref = 1\global\xdef\therefs{\item{[\number\nref]} #2\ }
\else
\global\xdef\oldrefs{\therefs}
\global\xdef\therefs{\oldrefs\vskip.1in\item{[\number\nref]} #2\ }%
\fi%
\global\advance\nref by 1
}

\def\listrefs{\vfill\eject\section{References}\therefs}


\newcount\nfoot
\global\nfoot = 1

\def\foot#1#2{\xdef #1{(\number\nfoot)}
\footnote{${}^{\number\nfoot}$}{\eightrm #2}
\global\advance\nfoot by 1
}


\newcount\nfig
\global\nfig = 1
\def\thefigs{} 

\def\figure#1#2{\xdef #1{(\number\nfig)}
\ifnum\nfig = 1\global\xdef\thefigs{\item{(\number\nfig)} #2\ }
\else
\global\xdef\oldfigs{\thefigs}
\global\xdef\thefigs{\oldfigs\vskip.1in\item{(\number\nfig)} #2\ }%
\fi%
\global\advance\nfig by 1 } 

\def\fig#1{\xdef #1{(\number\nfig)}
\global\advance\nfig by 1 } 


\newcount\cflag
\newcount\nequation
\global\nequation = 1
\def\eqlabel{(1)}

\def\nexteqno{\ifnum\cflag = 0
\global\advance\nequation by 1
\fi
\global\cflag = 0
\xdef\eqlabel{(\number\nequation)}}

\def\lasteqno{\global\advance\nequation by -1
\xdef\eqlabel{(\number\nequation)}}

\def\label#1{\xdef #1{(\number\nequation)}
\ifnum\dflag = 1
{\escapechar = -1
\xdef\draftname{\littlefont\string#1}}
\fi}

\def\clabel#1#2{\xdef\eqlabel{(\number\nequation #2)}
\global\cflag = 1
\xdef #1{\eqlabel}
\ifnum\dflag = 1
{\escapechar = -1
\xdef\draftname{\string#1}}
\fi}

\def\cclabel#1#2{\xdef\eqlabel{#2)}
\global\cflag = 1
\xdef #1{\eqlabel}
\ifnum\dflag = 1
{\escapechar = -1
\xdef\draftname{\string#1}}
\fi}


\def\eeq{}

\def\eqnn #1\eeq{$$ #1 $$}

\def\eq #1\eeq{
\ifnum\dflag = 0
{\xdef\draftname{\ }}
\fi 
$$ #1
\eqno{\eqlabel \rlap{\ \draftname}} $$
\nexteqno}







\def\eqa #1\eeq{
\ifnum\dflag = 0
{\xdef\draftname{\ }}
\fi 
$$ \eqalignno{ #1 } $$
\global\cflag = 0}


\def\etc{{\it etc.\/}}

\def\cf{{\it c.f.\/}}


\def\plb#1#2#3{{\it Phys.\ Lett.} {\bf #1B} (19#2) #3}

\def\prd#1#2#3{{\it Phys.\ Rev.} {\bf D#1} (19#2) #3}

\def\prl#1#2#3{{\it Phys.\ Rev.\ Lett.} {\bf #1} (19#2) #3}


\global\nulldelimiterspace = 0pt



\def\frac#1#2{{{#1} \over {#2}}\,}  
\def\hf{{1\over 2}}



\def\Dsl{\hbox{/\kern-.6700em\it D}} 
\def\dsl{\hbox{/\kern-.5300em$\partial$}}
\def\pxpsl{\hbox{/\kern-.5600em$p$}}
\def\ssl{\hbox{/\kern-.5300em$s$}}
\def\epssl{\hbox{/\kern-.5100em$\epsilon$}}
\def\delsl{\hbox{/\kern-.6300em$\nabla$}}
\def\lxpsl{\hbox{/\kern-.4300em$l$}}
\def\elxpsl{\hbox{/\kern-.4500em$\ell$}}
\def\kxpsl{\hbox{/\kern-.5100em$k$}}
\def\qxpsl{\hbox{/\kern-.5000em$q$}}
\def\transp#1{#1^{\sss T}} 
\def\sla#1{\raise.15ex\hbox{$/$}\kern-.57em #1}



\def\roughly#1{\mathrel{\raise.3ex
\hbox{$#1$\kern-.75em\lower1ex\hbox{$\sim$}}}}
\def\lsim{\roughly<}



\def\bfe{{\bf e}}
\def\bff{{\bf f}}

\def\bfm{{\bf m}}


\def\Bfi{{\bf I}}

\def\Bfv{{\bf V}}


\def\Scl{{\cal L}}


\def\ssh{{\sss H}}

\def\ssl{{\sss L}}

\def\ssn{{\sss N}}




\def\avg#1{\langle #1 \rangle}



\def\hc{{\rm h.c.}}


\def\eV{{\rm \ eV}}

\nopageonenumber
\baselineskip = 16pt
\barsoff


\def\veps{\varepsilon}
\def\eps{\epsilon}

\def\bk{\item{}}
\def\bb{\beta \beta}
\def\bbzn{\beta \beta_{0\nu}}
\def\nusol{\nu_\odot}
\def\nuatm{\nu_{\rm atm}}
\def\pp{$p$-$p$}
\def\bo{${}^8$B}
\def\be{${}^7$Be}
\def\nue{\nu_e}
\def\num{\nu_\mu}
\def\nut{\nu_\tau}
\def\nui{\nu_i}
\def\emu{$\nue - \num$}
\def\etau{$\nu_e - \nut$}
\def\mutau{$\num - \nut$}

\def\mnui{m_{\nui}}
\def\mnue{m_{\nue}}

\def\evsq{\eV^2}
\def\si{s_i}
\def\Oij{O_{ij}}
\def\cs#1{c_#1}
\def\sn#1{s_#1}
\def\csa{\cs{\alpha}}
\def\csb{\cs{\beta}}
\def\cst{\cs{\theta}}
\def\sna{\sn{\alpha}}
\def\snb{\sn{\beta}}
\def\snt{\sn{\theta}}


\rightline{February, 1994.}
\rightline{McGill-94/07, NEIP-94-003}
\rightline{hep-ph/9402229}
\title
\centerline{$\hbox{\titlefont Naturally Degenerate
Neutrinos}^\dagger$\footnote{}{${}^\dagger$
\eightrm Research partially supported by the Swiss National Foundation.}}
\endtitle
\centerline{P. Bamert and C.P. Burgess${}^{*}$\footnote{}{${}^*$ {\eightrm
Permanent Address: Physics Department, McGill University, 3600 University St.,
Montr\'eal, }}
\footnote{}{\eightrm $\phantom{{}^*}$ Qu\'ebec,  Canada, H3A 2T8.}}
\vskip .1in
\centerline{\it Institut de Physique, Universit\'e de Neuch\^atel}
\centerline{\it 1 Rue A.L. Breguet, CH-2000 Neuch\^atel, Switzerland.}
\vskip .3in
\abstract
The solar neutrino problem, atmospheric neutrino problem and the
hot dark matter which seems to be required by galaxy formation can all be most
economically described by three very degenerate neutrinos with a common mass
of a few eV. Such neutrino masses could also be relevant for the recent data on
double beta decay. We provide an existence proof for such a scenario by
constructing an explicit model in which the desired hierarchy of neutrino mass
splittings arises naturally.
\endabstract


\section{Introduction}

\ref\nuphys{Some surveys with references can be found in the contributions
of T.J. Bowles, Y. Totsuka, and J. Rich and M. Spiro, in {\it Relativistic
Astrophysics  and Particle Cosmology}, ed. by C.W. Akerlof and M.A. Srednicki,
Annals of the New York Academy of Sciences, Vol 688, (New York, 1993).}
There have been a number of potential indications for new physics in the
neutrino  sector over the past five years \nuphys, but of these only two have
(so far) survived further scrutiny by later generations of experiments.
These are:

\item
{\bf 1.} {\sl The Solar-Neutrino ($\nusol$) Deficit:}
The $\nusol$ flux, now observed with Cl, water and Ga detectors, remains
in conflict with the predictions of the Standard Solar Model (SSM).
The observations appear to be consistent with ($i$) no
suppression of the  dominant \pp\ cycle neutrinos, ($ii$) partial
suppression of the \bo\ neutrinos,  and ($iii$) virtually complete
elimination of the \be\ neutrinos.

\item
{\bf 2.} {\sl The Atmospheric-Neutrino ($\nuatm$) Problem:}
Both the Kamiokande and IMB detectors detect $\nuatm$'s, but with
a ratio of $\num/\nue$ which is roughly 60\% of what is theoretically expected
--- a result which is consistent with the null results of Fr\'ejus and which
more recently appears to be confirmed by SOUDAN-II measurements.

\ref\darkmatter{See, for example, the contributions by P.J.E. Peebles,
J.L. Tonry, and K. Griest, {\it ibid}.}
A third, more tentative, indication for new physics comes from cosmology and
astrophysics \darkmatter. There is now compelling evidence that the bulk of the
matter of the universe is invisible, and reasonably persuasive arguments that
much or most of this dark matter is not baryons. Furthermore,  the observed
distribution of matter on the largest scales, together with  the recently
measured pattern of fluctuations in the cosmic microwave  background, appear to
indicate that roughly 30\% of this particle dark  matter should have been
relativistic during the crucial epoch for galaxy formation. If this hot, dark
matter should consist of the known neutrinos, certainly the most conservative
assumption, then their masses should satisfy $\sum_i \mnui  \simeq 7 \, \eV$.

\ref\degenerates{D. Caldwell and R. Mohapatra, \prd{48}{93}{3259};\bk
S.T. Petcov and A.Yu. Smirnov, \plb{322}{94}{109};\bk
A. Joshipura, Ahmedabad preprint.}
As has recently been pointed out \degenerates, all three of these hints for
new physics in the neutrino sector can be very economically understood
in terms of a specific pattern of masses involving only the three known
neutrino species.  Moreover, for only three neutrinos, the required
pattern of masses and mixings is very tightly constrained by the requirement
that all three phenomena be incorporated. The conditions are:
\item
{\it i.} In order to account for the $\nuatm$ anomaly, one pair
(either \emu\ or \mutau) of neutrinos must  have large mixings,
$\sin^2 2\theta_{ij} \simeq 0.5$, with a squared-mass splitting: $\Delta
m^2_{ij} \equiv m^2_{\nu_i} - m^2_{\nu_j} \simeq 10^{-2} \, \evsq$.
\ref\enlarged{A. Joshipura and P. Krastev, preprint IFP-472-UNC,
PRL-TH-93/13, hep-ph/9308348.}
\item
{\it ii.} In order to provide an MSW explanation for the $\nusol$
deficit, another pair (either \emu\ or \etau)  of neutrinos must have
a squared-mass splitting: $\Delta m^2_{ij} \simeq 10^{-5} \, \evsq$
and either comparatively large mixings, $\sin^2 2\theta_{ij} \simeq
0.5$, or small mixings, $\sin^2 2\theta_{ij} \simeq 10^{-2}$. Furthermore,
if it is \emu\ oscillations that are the explanation for the $\nuatm$ anomaly,
and it is \etau\ that resonantly oscillates inside the sun, then {\it both}
of these oscillations deplete the solar neutrino flux, and the
phenomenologically acceptable range of \etau\ mass splittings is enlarged to
include $\Delta m^2_{e\tau} \simeq 10^{-4} \, \evsq$ \enlarged.
\item
{\it iii.} Finally, the three species of light neutrinos can make up the
required hot, dark matter provided their masses are degenerate, with
their common mass being given by $\mnui \simeq 2 \, \eV$.

\ref\bbznbound{A recent detailed review, with references to the
experiments is given by:\bk  M. Moe, {\it Int. J. Mod. Phys.} {\bf E2}  (1993)
507.}
It is noteworthy that $\mnue$ in this range would imply a $\bbzn$
signal that is just consistent with the present upper bound \bbznbound,
$\mnue \lsim 1$ eV, keeping in mind a possible uncertainty of a factor of
two in this bound due to difficulties in computing nuclear matrix elements.
This would be of particular interest should the presently observed
2-$\sigma$ bump at the electron endpoint persist in the $\bb$ data.

\ref\seesaw{M. Gell-Mann, P. Ramond and R. Slansky, in {\it Supergravity},
ed. by F. van Nieuwenhuizen and D. Freedman (Amsterdam, North Holland,
1979) 315; \bk
T. Yanagida, in the Proceedings of the Workshop on {\it Unified Theory and
Baryon Number in the Universe}, ed. by O. Sawada and A. Sugamoto (KEK,
Tsukuba, 1979) 95;\bk
R.N. Mohapatra and G. Senjanovi\'c, \prl{44}{80}{912}.}
The most striking feature about the required neutrino mass pattern is that it
is very degenerate, with the pairs of oscillating states being respectively
split by only $10^{-2}$ eV (and $10^{-4}- 10^{-5}$ eV) out of a total mass
which
is of order 1 eV. As was remarked in some of refs. \degenerates, such a
degenerate pattern is  difficult to come by in a natural way in renormalizable
gauge theories and, in particular, at first sight appears not to be what is
expected of a `see-saw' mechanism for generating neutrino masses.  A corollary
is that we stand to learn something interesting about the nature of the physics
which underlies these neutrino masses if this should prove to be how neutrino
masses are realized in nature.

It is the purpose of the present note to furnish an existence proof for
models which can naturally generate the required type of light-neutrino
masses and mixings. In fact, we do so within a see-saw framework in which the
light-neutrino masses are intimately related to those of a family of heavy,
sterile neutrinos. In particular, we imagine the existence of an approximate
family symmetry which mixes the light neutrinos amongst themselves, but
which is broken by the couplings of  some of the heavy, singlet neutrinos.
Then the required hierarchy of small mass splittings is obtained in terms
of heavy mass {\it ratios}, and do not require the introduction of new
dimensionless couplings that are much smaller than unity.

We organize our presentation in the following way. Section 2 describes
some general features of a light-neutrino mass matrix which satisfies all of
the
phenomenological constraints. We explicitly exhibit its masses and mixing
angles, and demonstrate that these include the parameter ranges that are
required. An explicit model which produces this mass matrix is then constructed
in Section 3. We finish with some concluding remarks in Section 4.

\vfill\eject
\section{A Desirable Mass Matrix}

The main observation for the purposes of model building is that the
light-neutrino mass matrix should be a sum of successively
smaller contributions,
\label\masssum
\eq
\bfm = \bfm_0 + \bfm_1 + \bfm_2 ,
\eeq
with $\bfm_0 = m_0 \, \Bfi$ proportional to the $3\times 3$ unit matrix,
$\bfm_1$ a rank-one matrix with elements which are $O(10^{-2} \, m_0)$
in magnitude, and $\bfm_2$ a generic matrix whose elements are
$O(10^{-4} \, m_0)$.

A representative example of a matrix of this type, and which is of the form
that
is actually generated by the models of the next section, is:
\label\repmatrix
\eq
\bfm = m_0 \Bigl[ \Bfi + \eps \; \bfe \, \transp{\bfe} - \delta \;
\bigl( \bfe \, \transp{\bff} + \bff \, \transp{\bfe} \bigr) + \xi \;
\bff \, \transp{\bff} \Bigr].
\eeq
In this expression, $m_0 \simeq 2$ eV sets the scale of the common
overall mass, $\bfe$ and $\bff$ are arbitrary unit vectors in neutrino
flavour-space, and $\transp{\bfe} \bfe \equiv \sum_{i=1}^3 e^2_i =1$
\etc\
$\eps$, $\delta$ and $\xi$ are three quantities which are all taken to be
much smaller than unity.

In what follows it is worth keeping in mind the sizes for $\eps$, $\delta$ and
$\xi$ that actually arise in the models considered in later section. There are
two cases of interest. In Case I we typically find $\eps \sim \veps^2 \gg
\delta
\sim \veps^3 \gg \xi \sim \veps^4$, where $\veps \lsim 0.1$ is a  smallish
parameter of the model. For case II, however, the sizes of $\eps$ and
$\xi$ are reversed, so that: $\eps \sim \veps^4 \ll \delta \sim \veps^3 \ll \xi
\sim \veps^2$.

It is straightforward to diagonalize the matrix of eq.~\repmatrix. Denoting
the cosine of the angle, $\alpha$, between the vectors $\bfe$ and $\bff$ by
$\csa$, we have $\bfe \cdot \bff = \csa$, and the eigenvalues of $\bfm$ are
given by:
\label\eigenvalues
\eq  \eqalign{
m_\pm &= m_0 \left\{ \left( 1 + {\eps + \xi \over 2} -  \delta \csa \right)
\pm \hf \left[ \eps^2 + \xi^2 + 4 \delta^2 - 4  \delta ( \eps +
\xi) \csa- 2 \eps \xi(1 - 2\csa^2)  \phantom{\hf} \right]^{1/2} \right\} \cr
\hbox{and:} \qquad m_3 &= m_0. \cr}
\eeq
The corresponding eigenvectors are:
\label\eigenvectors
\eq
{ \bfe_+  \choose \bfe_- } =  \pmatrix{ \cst
& \snt \cr - \snt & \cst \cr}  \;  { \bfe \choose \bfe' }
 , \qquad \hbox{and} \qquad \bfe_3
= \bfe \times \bfe',
\eeq
where $\bfe'$ is a unit vector which is defined to be orthogonal to $\bfe$ and
coplanar with $\bfe$ and $\bff$. It is given explicitly by: $\bfe' =
{1 \over \sna} \; ( \bff - \csa \; \bfe)$. Finally, the angle $\theta$ which
appears in the rotation matrix in eq.~\eigenvectors\ is given by:
\label\thetadef
\eq
\tan 2\theta = {2 (\xi \csa - \delta) \sna \over \eps - 2 \delta \csa
- \xi(1-2\csa^2)}.
\eeq

If we take the original mass matrix to be in a basis for which the charged
leptons are diagonal, then the relation between the weak-interaction
and the propagation eigenstates is:
\label\kmmatrix
\eq
\pmatrix{ \nue \cr \num \cr \nut \cr} =
\pmatrix{e_1 \cst + {\snt \over \sna} \, (f_1 - e_1 \csa) &
- e_1 \snt + {\cst \over \sna} \, (f_1 - e_1 \csa) &
{1\over \sna} \, (e_2 f_3 - e_3 f_2) \cr
e_2 \cst + {\snt \over \sna} \, (f_2 - e_2 \csa) &
- e_2 \snt + {\cst \over \sna} \, (f_2 - e_2 \csa) &
{1\over \sna} \, (e_3 f_1 - e_1 f_3) \cr
e_3 \cst + {\snt \over \sna} \, (f_3 - e_3 \csa) &
- e_3 \snt + {\cst \over \sna} \, (f_3 - e_3 \csa) &
{1\over \sna} \, (e_1 f_2 - e_2 f_1) \cr} \;
\pmatrix{ \nu_+ \cr \nu_- \cr \nu_3 \cr}
\eeq

Notice that for one of the cases of practical interest, where $\eps \sim
\veps^2 \gg \delta \sim \veps^3 \gg \xi \sim \veps^4$, we have the following
approximate expressions:
\label\approxevs
\eq  \eqalign{
m_+ &= m_0 \left[ 1 + \eps - 2 \csa \delta + \csa^2 \xi + {\delta^2
\sna^2 \over \eps}  + O(\veps^5) \right], \cr
m_- &= m_0 \left[ 1 +  \sna^2 \xi - {\delta^2 \sna^2  \over \eps}
+ O(\veps^5) \right], \cr
\tan 2\theta &\approx {-2 \delta \sna \over \eps} \sim O(\veps).
 \qquad  \qquad  \qquad \hbox{(Case I)} \cr}
\eeq
These expressions imply the mass splittings: $\Delta m^2_{+-} \sim
\Delta m^2_{3+} \sim O(\veps^2)$ and $\Delta m^2_{3-} \sim O(\veps^4)$.
Notice that the smaller splitting here is $O(\veps^4)$ even though the
parameter $\delta$ is itself of order $\veps^3$.

In the alternative limit, Case II, we have approximate eigenvalue expressions
which are simply obtained from eq.~\approxevs\ by making the replacement $\eps
\leftrightarrow \xi$, and so which give mass splittings which are the same
order
in $\veps$ as for Case I. By contrast, the mixing angle, $\theta$, is now given
approximately by:
\label\approxmix
\eq
\tan 2\theta \approx {-2 \csa \sna \over 1 - 2 \csa^2} = \tan 2\alpha \sim
O(1) \qquad \hbox{(Case II)}.
\eeq

Before turning to model building, we first pause to explicitly display two
representative sets of parameters, corresponding to the two possible
neutrino-mixing scenarios. Firstly, for \emu\ atmospheric-neutrino
oscillations,
and \etau\ MSW mixing we take Case I, for which $\eps \simeq 10^{-2}$, $\delta
\simeq 10^{-3}$, $\xi \simeq 10^{-4}$ and $\theta \simeq 0.1$. In this case if
we choose $e_3 = f_1 = f_2 = 0$, $f_3 = 1$, $e_1 = \snb$ and $e_2 = \csb$, for
some angle $\beta$, then $\csa = 0$ and the mixing matrix of eq.~\kmmatrix\
simplifies to: %
\label\scenarioI
\eq
\Bfv = \pmatrix{\snb \cst & - \snb \snt & \csb \cr
\csb \cst & - \csb \snt & - \snb \cr \snt & \cst & 0 \cr}.
\eeq
We see that the atmospheric oscillations are, in this scenario, controlled
$\sin^2 2\beta \simeq 0.5$, and that MSW oscillations are governed by $\sin^2
2\theta \simeq 10^{-2}$. Notice that this last choice follows from the
general result, $\theta \sim O(\veps)$, for Case I.

Alternatively, if we wish to describe atmospheric neutrinos with
\mutau\ oscillations, and the MSW effect occurs with \emu\ oscillations we work
with Case II. Then $\eps \simeq 10^{-4}$, $\delta \simeq 10^{-3}$, $\xi
\simeq 10^{-2}$ and $\theta = \alpha \sim O(1)$. We choose $e_3 = 0$, $e_2 =
\csb$, $e_1 = \snb$, $f_3 = \sna$, $f_2 = \csa \csb$ and $f_1 = \csa \snb$. In
this case we again obtain the same mixing matrix as in eq.~\scenarioI, but with
$\theta = \alpha$ not required to be small. The large-angle \mutau\
oscillations are now achieved
by choosing $\sin^2 2\alpha \simeq 0.5$, and the MSW oscillations are produced
by taking $\sin^2 2\beta \simeq 10^{-2}$.

\section{An Underlying Model}

We now require a model which naturally produces a light-neutrino mass matrix as
in eq.~\repmatrix. In order to do so we must ensure, as a first approximation,
a
degenerate set of eV neutrinos. We can easily do so by demanding an approximate
symmetry under which the light neutrinos rotate into one another in a real
representation. We therefore require an approximate $O(3)$ symmetry under which
the usual standard-model (SM) neutrinos, $\nu_i$, transform according to $\nu_i
\to \Oij \; \nu_j$, with $\Oij$ a real, 3 $\times$ 3, orthogonal matrix. This
symmetry can only be approximate since it is explicitly broken in the
standard model by the charged-current weak interactions (or, equivalently, the
charged lepton masses). An eV-size mass for these neutrinos is then obtained by
introducing three electroweak-singlet neutrinos, $\si$, which also transform as
triplets under the approximate $O(3)$ symmetry.

\ref\loopmasses{For a recent discussion see: D. Choudhury, R. Gandhi, J. Gracey
and B. Mukhopadhyaya, preprint CTP-TAMU-76/93, MRI-PHY-/12/93, LTH-326.}
With this particle content, the most general renormalizable and
$O(3)$-invariant
Yukawa couplings and mass terms (beyond the usual SM ones) are:
\label\invterms
\eq
\Scl_{\rm inv} = - {M \over 2} \; (\si \, \si) - \lambda \; (L_i \, \si) \; H +
\hc ,
\eeq
where $H = {\phi^+ \choose \phi^0}$ and $L_i = {\nui \choose e_i}$ are
respectively the SM Higgs and lepton doublets. If $v = \avg{H} \simeq 174$ GeV
is the SM Higgs {\it vev}, and if $\lambda v \ll M$, then the light neutrinos
acquire degenerate masses whose size is given by $m_0 = \lambda^2 v^2/M$.
Choosing $m_0 \simeq 2$ eV then requires $M/\lambda^2 \simeq 2\times
10^{13}$ GeV.

For the theory as stated so far, the three light neutrinos are not exactly
degenerate because electroweak radiative corrections can split them. At one
loop this splitting is proportional to the corresponding charged-lepton masses,
although this need not be so for higher loops \loopmasses. Unfortunately, these
radiatively-generated splittings are too small to account for the required mass
pattern and so we must introduce a further source of $O(3)$ symmetry breaking.
In order to produce the desired mass matrix of eq.~\repmatrix, we require
$O(3)$-breaking order parameters, $e_i$ and $f_i$, which transform as triplets.
We therefore supplement the model with one more singlet neutrino, $N$, which is
also neutral under $O(3)$. We permit the couplings of $N$ to the other fields
to
explicitly break the $O(3)$ symmetry. The most general such renormalizable
couplings for the $N$ field are then
\label\ncouplings
\eq
\Scl_{\ssn} = - {m \over 2} \; (N \, N) - \mu_i \; (\si \, N) -
g_i \; (L_i \, N) \; H + \hc
\eeq
We return to the question of how natural it is to choose these particular
symmetry-breaking couplings below.

For $\lambda v, g_i v \ll M, m, \mu_i$, the complete mass matrix for the light
neutrinos that is generated at tree level by the interactions of
eqs.~\invterms\
and \ncouplings\ has the form of eq.~\repmatrix, with
\label\mparams
\eq
m_0 = {\lambda^2 v^2 \over M}, \qquad \eps = {\mu^2 \over m M} , \qquad
\delta = {\mu g \over m \lambda},  \qquad \xi = {g^2 M \over
\lambda^2 m}, \qquad e_i = {\mu_i \over \mu}, \qquad f_i = {g_i \over g}
\eeq
where
\label\newdefs
\eq
\mu^2 \equiv \sum_i \mu^2_i \qquad \hbox{and} \qquad g^2 \equiv \sum_i g^2_i .
\eeq

Representative values for model parameters which reproduce Cases I and II of
the
previous sections are easily found. For example, Case I is reproduced by any of
the following choices ($\veps = 0.1$): (Ia) $m \sim M$, $\mu_i \sim \veps M$,
$g_i \sim \veps^2 \lambda$; (Ib) $\mu_i \sim M \sim \veps^2 m$, $g_i \sim \veps
\lambda$; (Ic) $M \sim \veps^4 m$, $\mu_i \sim \veps^3 m$, $g_i \sim \lambda$.
Similarly, Case II is reproduced by any of the following: (IIa) $M \sim \veps^2
m$, $\mu_i \sim \veps^3 m$, $g_i \sim \lambda$; (IIb) $\mu_i \sim \veps^2 M$,
$M \sim m$, $g_i \sim \veps \lambda$; (IIc) $\mu_i \sim M \sim \veps^4 m$,
$\lambda \sim \veps g_i$.

We finally consider the issue of naturalness. In this model, the required small
splittings amongst the light-neutrino masses are ultimately due to the
relative size of the symmetry breaking terms, $\mu_i$ and $g_i$, in
comparison with the $O(3)$-invariant couplings, $M$, $m$ and $\lambda$. Notice
however that the ratios of these couplings are actually not required to be
extremely small, since they may all be chosen to be no smaller than $\veps^2
\simeq 0.01$, say (\cf\ examples (Ia), (Ib) and (IIb) above). This is more than
adequate for producing the desired mass pattern.

The final question concerns loop corrections. Is it consistent to choose the
$O(3)$-breaking interactions as large as we have and yet {\it not} to include
any $N$-independent $O(3)$-breaking terms, such as $(s_i \, s_j)$, $(L_i \,
s_j) H$ or $(L_i L_j) H H$? The answer to this question is a resounding
`yes', although this is subject to  certain weak constraints on the various
couplings of the model, which we now describe.

It is clear that loops involving
the symmetry-breaking $N$-interactions must inevitably induce all possible
$O(3)$-breaking couplings amongst the other neutrinos. But the point is that
unless $g_i$ or $g_i/\lambda$ should be much larger than the range 0.1 to
0.001,
these induced
terms only perturb the above mass eigenvalues by amounts that are smaller than
a
part in  $10^4 - 10^5$, and so they are negligible in comparison to the effects
which we have considered. This is particularly clear for the induced
contributions to symmetry-breaking operators like $(s_i \, s_j)$, since the
one-loop contributions to these are strongly suppressed by small mass
insertions
or loop factors. (These suppressions arise because any symmetry-breaking loop
must necessarily involve a virtual $N$ field, and this only becomes possible,
given only two external $\si$ lines, at two loops and beyond, or at one loop
with a number of mass insertions.)

Loop-induced symmetry-breaking operators of the form $(L_i \, s_j)  H$ and
$(L_i L_j) H H$ {\it can} arise at one loop, typically being generated through
flavour-changing contributions to the various neutrino kinetic terms. Their
generic size is respectively of order $\lambda g_i g_j/(16 \pi^2)$ and $g_i
g_j/(16 \pi^2)$, which is innocuous provided that $g_i$ and
$g_i/\lambda$ are smaller than $O(0.1)$. (This therefore excludes examples
Ic and IIa above.) The most dangerous of these contributions are those which
are
also enhanced by large logarithms, such as the one-loop contribution to the
operator $(L_i L_j) H H$ which is induced by the Higgs self-coupling $\zeta \,
(H^\dagger H)^2$. This graph contributes an amount to $\xi$ which is
$\zeta$-dependent, but of order $\zeta
(g/4\pi \lambda)^2 \, \ln (m_\ssh^2/ M^2)$. For $\zeta \sim 0.1$ and for
small Higgs mass, $m_\ssh \sim 80$ GeV, this can be numerically as large as
$\sim 0.1 (g/\lambda)^2$. Such an enhancement can strengthen the bound
on $g/\lambda$ by factors of 3 or so.

\section{Concluding Remarks}

In summary, by constructing an explicit underlying model we provide an
existence proof for the recently-proposed economical scenario in which
the solar- and atmospheric-neutrino data, and the existence of hot dark
matter are all explained in terms of the masses and mixings of the three
known light neutrinos. This picture is attractive because it is tightly
constrained, and so is quite predictive. In particular, it implies a $\bbzn$
signal at the current experimental sensitivity.

The underlying model realizes this mass pattern within a see-saw type
framework in which the light neutrinos mix with a family of heavier,
sterile neutrinos. The degeneracies of the light neutrino masses emerge
in this picture as the low-energy residue of an approximate $O(3)$
neutrino-flavour symmetry, which is broken only by the interactions of
one sector of the heavy neutrinos. All of the desired small mass splittings
amongst the light neutrinos then naturally arise as ratios of the
sterile-neutrino masses, without requiring the introduction of small
dimensionless Yukawa couplings.

\bigskip

\centerline{\bf Acknowledgments}

\bigskip

C.B. would like to thank the organizers and participants of the WHEPP III
workshop in Madras, India, for providing the milieu which stimulated this
research, and in particular A. Joshipura for sharing an early version of
ref.~\degenerates.  We also acknowledge partial funding by N.S.E.R.C.\ of
Canada, les Fonds F.C.A.R.\ du Qu\'ebec, and the Swiss National Foundation.

\bigskip\bigskip

\ref\latecomers{D. Caldwell and R. Mohapatra, preprint UCSB-HEP-94-03,
UMD-PP-94-90, hep-ph/9402231; \bk
A. Ioannissyan and J.W.F. Valle, preprint FTUV/94-08, IFIC/94-05,
hep-ph/9402333;\bk
D. Lee and R. Mohapatra, preprint UMD-PP-94-95, hep-ph/9403201;\bk
A. Joshipura, Ahmedabad preprint.}
\noindent{\it Note Added:} Shortly after completing this paper we received
several others which approach the same problem from the point of view of
$SO(10)$
grand-unified theories \latecomers.

\listrefs

\bye